\documentclass[pdflatex,sn-nature]{sn-jnl}

\usepackage{graphicx}
\usepackage{amsmath,amssymb,amsfonts}
\usepackage{booktabs}
\usepackage{array}
\usepackage{multirow}
\usepackage{xcolor}
\usepackage{lineno}

\begin{document}

\title[Calibration of tabular foundation models]{Tabular foundation models are not a shortcut to calibrated clinical risk prediction: a pre-registered simulation against the true risk}

\author*[1]{\fnm{Ebrahim Khaled} \sur{Ebrahim}}\email{ebrahimkhaled@alexu.edu.eg}

\affil*[1]{\orgdiv{Department of Applied Statistics, Faculty of Business}, \orgname{Alexandria University}, \orgaddress{\city{Alexandria}, \country{Egypt}}}

\abstract{Tabular foundation models are pretrained networks that predict risk from a table without fitting a model. Whether their risks are correct (calibrated) has been judged on real data, where true risk is unknown. In a pre-registered simulation with known true risks, we compared TabPFN-3.5, TabICLv2 and Mitra-v2 with Firth logistic regression with intercept correction (FLIC) in 58 scenarios and 1,000 random designs, then on four clinical data sets. Foundation models were not a shortcut to calibration. Their calibration slope matched FLIC's within the pre-registered margin in 36 or 37 of 48 core scenarios and was better only at $n=200$ with a 3\% event rate. Even 3,000 patients were often not enough: with FLIC, TabPFN-3.5 or Mitra-v2, 63\%--83\% of samples gave risks close to the truth in average, intercept and slope at event rates of 10\% and 30\%, and 35\%--50\% at 3\%. TabICLv2 did worse (8\%--39\% and 1\%--3\%). Across the random designs, foundation models had a lower true calibration error than FLIC beyond the margin in 25\%--38\% and a higher one in none, though small losses were predicted at $n=3000$. On real data, no model resisted a change in event rate between populations; recalibration on 500 local patients largely repaired it.}


\date{}
\maketitle

\section{Introduction}\label{sec:intro}

A clinical prediction model turns a patient's characteristics into a probability of an outcome, such as death within 28 days, on which treatment decisions may rest. The probability itself must then be right, not only the ranking of patients. A model is well calibrated when its predicted risks agree with the actual risks: among patients given a risk of 20\%, about 20\% have the event\cite{vancalster2019}. A poorly calibrated model can mislead even when it ranks patients well, because it overstates or understates the risk on which a decision rests\cite{vancalster2019}. Reporting guidance therefore asks every clinical prediction model to report its calibration\cite{collins2024,vancalster2025measures}.

Tabular foundation models (TFMs) offer a new way to build such a model. A TFM is a large neural network (a transformer) pretrained on millions of simulated data tables. It reads the development data as a prompt and returns a risk for each new patient in one pass, without fitting parameters to those data. TabPFN was the first widely used TFM\cite{hollmann2023,hollmann2025}. Its newest version, TabPFN-3.5\cite{jaeger2026tabpfn35}, and two other current models rank at or near the top of public benchmarks of prediction on tables\cite{erickson2025tabarena,grinsztajn2026tabpfn3}: TabICLv2, an open model designed to scale to large tables\cite{qu2025tabicl,qu2026tabiclv2}, and Mitra-v2, pretrained on a mixture of synthetic data generators\cite{zhang2025mitra,tao2026mitra}. They need no tuning and do well on small samples, which suits clinical prediction. A systematic review found that earlier machine-learning methods gave no performance benefit over logistic regression for clinical prediction\cite{christodoulou2019}, but early clinical studies of TabPFN have reported discrimination equal to or better than that of logistic regression and gradient boosting\cite{varghese2026,wu2026,shaktah2026}.

Whether TFM risks are calibrated is harder to answer. On real data the true risk of a patient is unknown; only the outcome, 0 or 1, is observed. Calibration must then be estimated by grouping or smoothing the outcomes against the predictions, and these estimates are noisy. The expected calibration error (ECE), common in machine learning, averages the gap between predicted and observed risk over bins of predictions and is biased in finite samples\cite{roelofs2022}. The integrated calibration index (ICI), common in medical statistics, is the mean absolute gap between each prediction and a smoothed curve of the outcomes\cite{austin2019}; it needs a large validation sample to be precise\cite{riley2021}. Calibration also depends on the development sample: a model developed on few patients tends to give risks that are too extreme\cite{riley2020,riley2025ai}.

The evaluations of TFM calibration we found all used real data. Clinical studies reported the Brier score or calibration plots; they found TabPFN well calibrated in some settings\cite{wu2026,varghese2026} and less so when the outcome was rare\cite{cui2026}. Machine-learning benchmarks reported the ECE, the log-loss or the coverage of prediction sets. They found that the TFMs with the best discrimination were not always the most reliable\cite{demelocosta2026,zhao2026tabrelibench}, that TabPFN ranked well on the ECE when applied to embeddings of images, text and audio\cite{zhang2026tabpfnembed}, and that recalibration after fitting can improve the probabilities of many classifiers, TFMs included\cite{berta2026calarena}. To our knowledge, none of these studies compared predictions with the true risk, none registered its analysis in advance, and none asked how large a development sample a TFM needs to be calibrated, the question that sample-size rules answer for regression models\cite{riley2020}.

We evaluate the calibration of TabPFN-3.5, TabICLv2 and Mitra-v2 in a pre-registered simulation study in which the true risk of every patient is known, so each prediction is compared with the true risk rather than with a noisy outcome. We compare the TFMs with five established methods. Maximum-likelihood logistic regression is the standard model. Firth logistic regression with intercept correction (FLIC), the reference, is a penalized logistic regression that stays stable with few events and makes the mean predicted risk equal the observed event rate\cite{puhr2017}. Ridge logistic regression shrinks the coefficients towards zero\cite{friedman2010}. A generalized additive model gives each continuous predictor a smooth non-linear effect\cite{wood2011}. LightGBM is an ensemble of boosted decision trees\cite{ke2017}, whose scores we recalibrate by Platt scaling, a logistic regression of the outcome on the scores. The protocol, including every rule that turns an estimate into a verdict, was registered before the main run\cite{osf2026}. Beyond a grid of 58 scenarios, 1,000 randomly drawn designs test whether the conclusions depend on the scenarios chosen, and four real clinical data sets, with a plasmode tier that places a known true risk on real patients' characteristics, give an exploratory check in real data.

We ask six questions: (Q1) how far from 1 each TFM's calibration slope lies at small development sizes, and in which direction (too extreme or too modest risks); (Q2) what development size each TFM needs to be calibrated, compared with FLIC; (Q3) how far the overall level of predicted risk is from the true level (calibration-in-the-large) when events are rare; (Q4) whether goodness-of-fit tests keep their nominal false-positive rate on TFM predictions, and how often they detect miscalibration; (Q5) how well recalibration on a small local sample repairs each TFM, and whether a test-guided rule picks the right cases; and (Q6) whether the findings from the fixed scenarios carry over to randomly drawn designs.

\section{Results}\label{sec:results}

Each simulated scenario fixes an outcome shape, an event rate $\pi$ (the share of patients who have the event) and a development size $n$. The four shapes are S1, linear on the logit scale; S2, smoothly non-linear; S3, with interactions; and S4, with an asymmetric link. The 48 core scenarios cross the four shapes with $\pi$ of 3\%, 10\% and 30\% and $n$ of 200, 500, 1,000 and 3,000; ten stress scenarios vary the predictors and the non-linearity. The primary measure is the true ICI, the mean absolute distance between each predicted risk and a smoothed curve of the true risk against the prediction (0 is perfect). A logistic regression of the true risk on the logit of the prediction gives two further measures. The true calibration slope $b^*$ is ideally 1; below 1 the predictions are over-confident (too extreme), above 1 under-confident (too modest). The true calibration intercept $a^*$ is ideally 0 and shows whether risks are too high or too low on average. An oracle that predicts the true risk itself has $a^*=0$ and $b^*=1$ exactly; its true ICI is small but not zero and shows the floor of the ICI estimator. Every model (arm) receives the same simulated data sets (replicates). Each TFM is compared with FLIC on the replicates common to both, and every estimate carries a Monte Carlo standard error (MCSE). A comparison is labelled \emph{equivalent} when its 95\% interval lies within a margin fixed before the run, \emph{better} or \emph{worse} when the interval lies wholly beyond the margin, and \emph{inconclusive} otherwise. The margins are wide: in 12 of the 48 core scenarios even the oracle, a perfect model, was labelled only equivalent to FLIC in true ICI (17 for $|\log b^*|$).

Each TFM failed on one of its 10,750 development sets, and ridge regression and LightGBM failed often enough at $\pi=3\%$ and $n=200$ that their results there are not labelled; all failures and degenerate fits are counted in Supplementary Table~S5. The oracle's true ICI lay between 0.0008 and 0.0041. Table~\ref{tab:sof} gives every pre-registered label.

\begin{table}[t]
\caption{\textbf{Summary of findings.} Counts of cells in each pre-registered label, for each TFM paired with FLIC on the replicates common to both. Labels come from 95\% intervals and margins fixed before the run: true ICI $\pm0.01$ ($\pm0.003$ at $\pi=3\%$), $|\log b^*|$ $\pm\log1.1$, $|a^*|$ $\pm0.05$. Eq., equivalent; inc., inconclusive. Q2 compares the smallest $n$ at which 80\% of replicates are weakly calibrated; an ordering is ``sensitive to Monte Carlo error'' when it changes if the lower 95\% limit of each share replaces the share. Q4a, size of the goodness-of-fit tests; Q4b, detection; Q5a, true ICI after logistic recalibration on 500 rows, against FLIC after the same recalibration and against the best arm (the lowest mean true ICI among FLIC and the TFMs after this recalibration and the other established methods as fitted); Q5b, positive predictive value of the test-guided rule. Shipped softmax temperature: TabPFN-3.5 1.0, TabICLv2 0.9; Mitra-v2 has none.}\label{tab:sof}
\footnotesize
\begin{tabular}{@{}>{\raggedright\arraybackslash}p{0.285\textwidth}>{\raggedright\arraybackslash}p{0.205\textwidth}>{\raggedright\arraybackslash}p{0.205\textwidth}>{\raggedright\arraybackslash}p{0.205\textwidth}@{}}
\toprule
Question (cells) & TabPFN-3.5 & TabICLv2 & Mitra-v2\\
\midrule
True ICI against FLIC (48 core scenarios; primary) & 44 eq., 4 better & 40 eq., 4 better, 4 worse & 44 eq., 2 better, 2 inc.\\
Q1 closeness of $b^*$ to 1 (8 main cells) & 5 eq., 3 inc. & 4 eq., 4 inc. & 4 eq., 4 inc.\\
Q1 closeness of $b^*$ to 1 (48 core scenarios) & 37 eq., 4 better, 7 inc. & 36 eq., 4 better, 8 inc. & 36 eq., 4 better, 8 inc.\\
Q1 direction of $b^*$ (8 main cells) & 2 close to 1, 1 under-confident, 5 unclear & 4 over-confident, 4 unclear & 4 close to 1, 4 unclear\\
Q2 minimum $n$ for weak calibration (12 shape--event-rate cells) & 9 the same $n$, 1 smaller, 2 larger & 10 the same $n$, 2 larger & 10 the same $n$, 2 larger\\
Q2 clarity of the ordering & 9 clear, 3 sensitive to Monte Carlo error & 10 clear, 2 sensitive & 10 clear, 2 sensitive\\
Q3 $|a^*|$ at $\pi=3\%$ (16 cells) & 12 eq., 1 better, 3 inc. & 4 eq., 3 better, 5 worse, 4 inc. & 12 eq., 4 inc.\\
Q4a size (11 tests) & 11 hold their size & 10 hold, 1 uneven & 10 hold, 1 uneven\\
Q4b EDGE minus Hosmer-Lemeshow (4 true-ICI points; all arms) & \multicolumn{3}{>{\raggedright\arraybackslash}p{0.6\textwidth}}{EDGE detects more often at 4 of 4 points (one comparison, pooled over arms)}\\
Q5a after logistic recalibration, against FLIC (58 scenarios) & 58 eq. & 58 eq. & 57 eq.\\
Q5a after logistic recalibration, against the best arm (48) & 30 within 0.005, 8 more than 0.005 above, 10 unclear & 30 within, 8 above, 10 unclear & 30 within, 8 above, 10 unclear\\
Q5b positive predictive value of the test-guided rule & at least 0.8 & at least 0.8 & at least 0.8\\
Q6 transfer of the grid to random designs & does not transfer (12 of 48 agree) & does not transfer (12 of 48) & does not transfer (15 of 48)\\
\botrule
\end{tabular}
\par\smallskip
\raggedright\footnotesize Q2 uses the replicates common to each TFM and FLIC, so two FLIC cells differ from Fig.~\ref{fig:minn}a, which uses all of FLIC's replicates: at $\pi=30\%$, FLIC's minimum $n$ is above 3,000 in S1 and 3,000 in S3 in the paired comparison. The size labels are defined in Methods; the size of every test on every arm, including FLIC, the oracle and the temperature arms, is in Supplementary Table~S9. For Q5b, FLIC was also labelled ``at least 0.8''.
\end{table}

\subsection{Distance from the true risk}\label{sec:res-ici}

\begin{figure}[t]
\centering
\includegraphics[width=\textwidth]{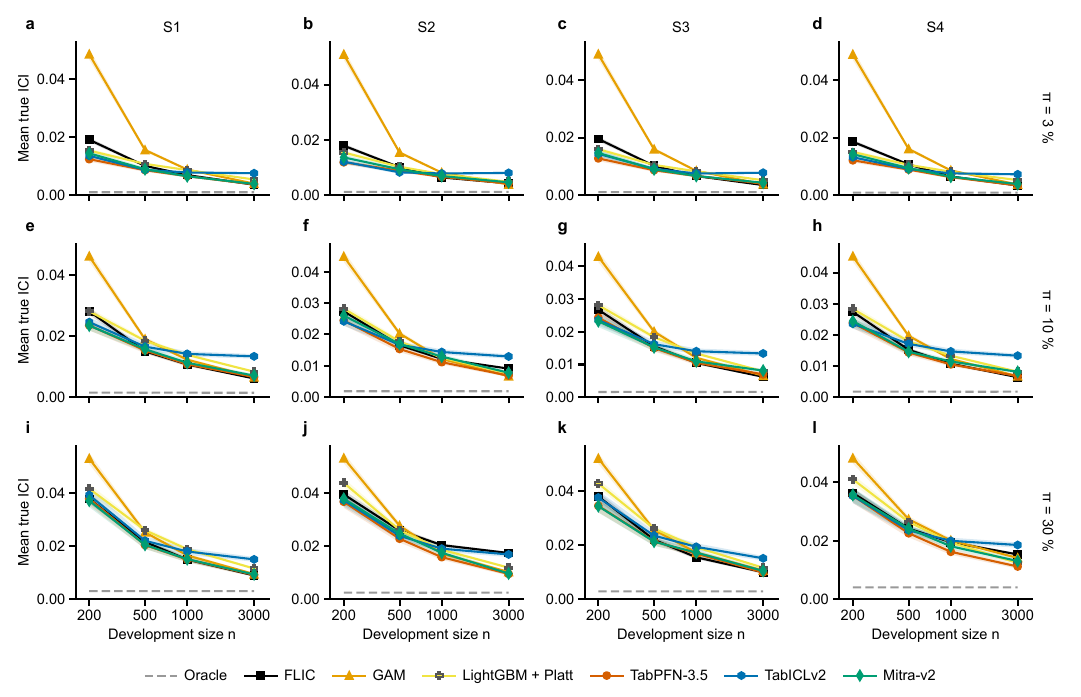}
\caption{\textbf{Distance of each model's predictions from the true risk.} Mean true integrated calibration index (ICI) against development size $n$, for event rates of 3\%, 10\% and 30\% (rows) and the four outcome shapes (columns: S1 linear, S2 smooth non-linear, S3 interactions, S4 asymmetric link). Points are means over replicates with $\pm1.96$ Monte Carlo standard errors (often smaller than the symbols). The dashed oracle line is the floor of the estimator. GAM, generalized additive model; LightGBM + Platt, gradient-boosted trees recalibrated by Platt scaling. Maximum-likelihood logistic regression and ridge regression are shown in Supplementary Information.}\label{fig:ici}
\end{figure}

TabPFN-3.5 and Mitra-v2 tracked FLIC in mean true ICI at every development size, event rate and shape (Fig.~\ref{fig:ici}). Paired with FLIC, TabPFN-3.5 was equivalent in 44 of the 48 core scenarios and better in 4; Mitra-v2 was equivalent in 44, better in 2 and inconclusive in 2. All the cells labelled better were at $\pi=3\%$ and $n=200$ (TabPFN-3.5: differences $-0.0053$ to $-0.0075$). TabICLv2 was equivalent in 40 scenarios, better in the same 4 cells and worse in all 4 cells at $\pi=3\%$ with $n=3000$ (differences 0.0036 to 0.0043); at $n=200$ it ran below its documented minimum of 300 training rows. Its mean true ICI reached a plateau: at $\pi=10\%$ it was 0.0140--0.0147 at $n=1000$ and 0.0130--0.0134 at $n=3000$, against 0.0061--0.0091 for FLIC and 0.0066--0.0070 for TabPFN-3.5 at $n=3000$. The generalized additive model was the worst arm at $n=200$ (0.043--0.053 across the twelve shape and event-rate cells, against 0.018--0.040 for FLIC).

\subsection{Calibration slope at small development size (Q1)}\label{sec:res-q1}

Q1 compares the distance of the slope from 1, $|\log b^*|$, with a margin of $\log 1.1$, in eight main cells (the four shapes at $\pi=10\%$, $n$ of 200 and 500). TabPFN-3.5 was equivalent to FLIC in 5 cells and inconclusive in 3; TabICLv2 and Mitra-v2 were each equivalent in 4 and inconclusive in 4 (Table~\ref{tab:sof}). The inconclusive cells were all at $n=200$, where every estimate of the paired difference (the Hodges-Lehmann estimate, a robust centre of the differences) favoured the TFM but the interval reached past the margin.

The direction differed between models (Supplementary Fig.~S1). FLIC was over-confident in 7 of the 8 main cells (median $b^*$ 0.720--0.752 at $n=200$). TabICLv2 was over-confident in the 4 cells at $n=200$ (median $b^*$ 0.825--0.853) and unclear in the other 4. Mitra-v2 was close to 1 in 4 cells and unclear in 4. TabPFN-3.5 leaned the other way (medians 1.023--1.186; under-confident in 1 cell). Across all 48 core scenarios, each TFM was equivalent to FLIC in 36 or 37 and better in the 4 cells at $\pi=3\%$ and $n=200$, where FLIC was strongly over-confident (median $b^*$ 0.528 in S1).

\subsection{Development size needed for calibration (Q2)}\label{sec:res-q2}

A fitted model is weakly calibrated when its mean risk is within 0.01 of the mean true risk and $|a^*|$ and $|b^*-1|$ are both at most 0.1, and moderately calibrated when, in addition, its true ICI is at most 0.015 (the 0.01 and 0.015 thresholds $\times0.3$ at $\pi=3\%$). A model's minimum $n$ is the smallest $n$ at which 80\% of its replicates reach a class. TabPFN-3.5 needed the same minimum $n$ for weak calibration as FLIC in 9 of the 12 shape and event-rate cells, a smaller $n$ in 1 and a larger $n$ in 2; TabICLv2 and Mitra-v2 needed the same $n$ in 10 and a larger $n$ in 2 (Table~\ref{tab:sof}). Every ``same $n$'' label means that neither model reached the 80\% rule by $n=3000$: a shared failure, not a shared success. Every ``smaller'' or ``larger'' label was sensitive to Monte Carlo error. With the thresholds doubled, TabICLv2 needed a larger $n$ than FLIC in all 12 cells, TabPFN-3.5 and Mitra-v2 in 2, and no TFM a smaller one.

\begin{figure}[t]
\centering
\includegraphics[width=\textwidth]{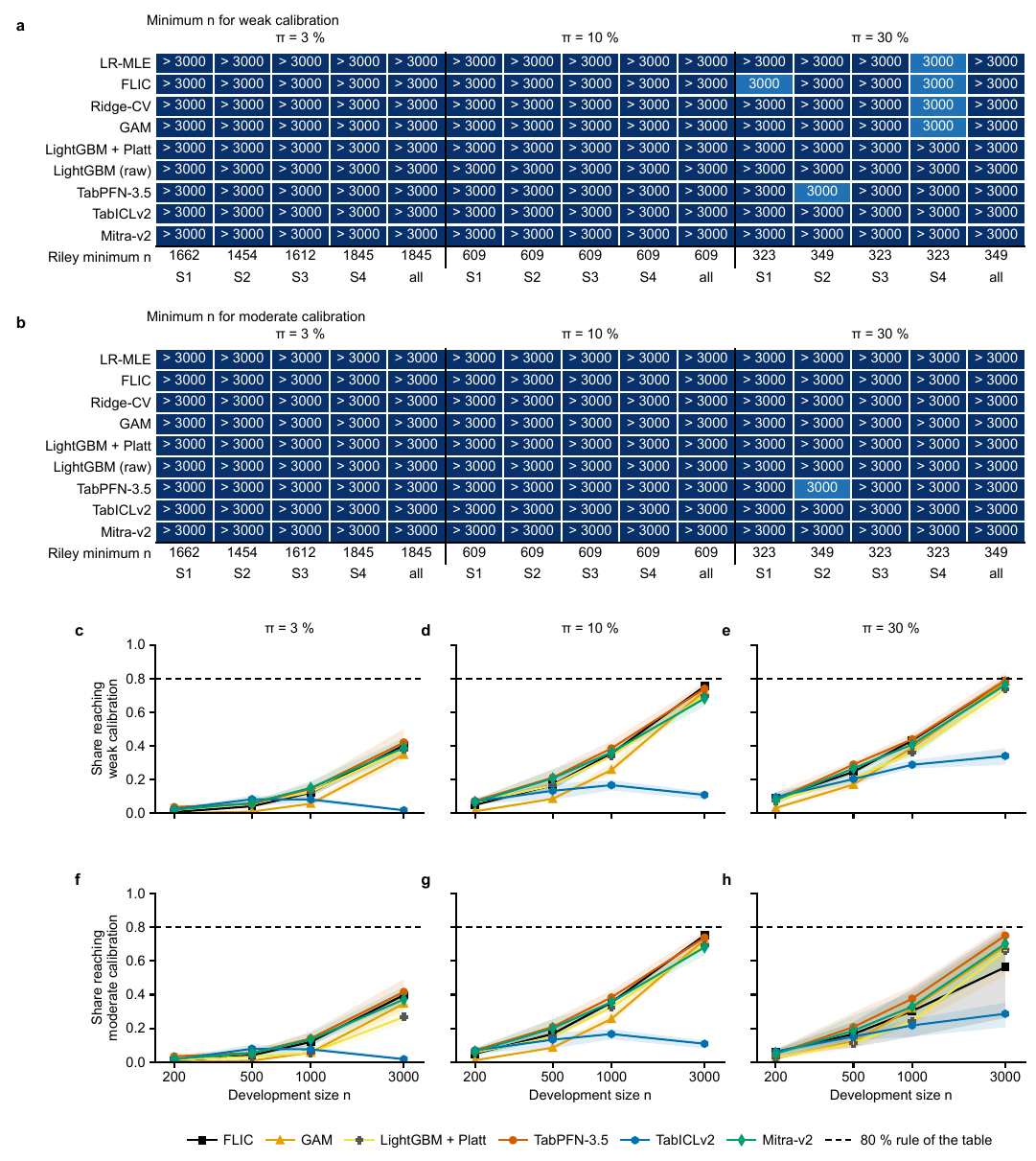}
\caption{\textbf{Development size needed for calibration.} \textbf{a},\textbf{b}, The smallest $n$ in the design at which each model reaches weak (\textbf{a}) or moderate (\textbf{b}) calibration in at least 80\% of replicates, by event rate and outcome shape (``$>3000$'' when no $n$ qualifies); the column ``all'' takes the largest value over the four shapes, and Riley's minimum sample size is given below each column. \textbf{c}--\textbf{h}, Share of development sets on which each model is weakly (\textbf{c}--\textbf{e}) or moderately (\textbf{f}--\textbf{h}) calibrated, against $n$; lines are the mean over the four outcome shapes and bands their range, and the dashed line marks the 80\% rule of panels \textbf{a} and \textbf{b}. The MCSE of each share is at most 0.022 for the established methods and 0.041 for the TFMs. LR-MLE, maximum-likelihood logistic regression; Ridge-CV, ridge logistic regression with the penalty chosen by cross-validation; GAM, generalized additive model; LightGBM + Platt and LightGBM (raw), gradient-boosted trees with and without Platt scaling. Versions at half and double the thresholds, with the lower 95\% limit of the share, and with the log-scale ICI are in Supplementary Information.}\label{fig:minn}
\end{figure}

The minimum-$n$ table shows why (Fig.~\ref{fig:minn}a,b). At $\pi=3\%$ and $10\%$, no model other than the oracle reached weak calibration in 80\% of replicates at any $n$ up to 3,000, in any shape. At $\pi=30\%$ and $n=3000$, six model--shape cells reached it, all close to the 80\% line: four were classical models in S4, FLIC also reached it in S1 (share 0.800; not on the replicates shared with the TFMs, Table~\ref{tab:sof}), and the only TFM was TabPFN-3.5 in S2. For moderate calibration, only TabPFN-3.5 in S2 reached the rule.

The shares separate the models further (Fig.~\ref{fig:minn}c--h). At $n=3000$ the share of weakly calibrated replicates for FLIC, TabPFN-3.5 and Mitra-v2 was 0.633--0.780 at $\pi=10\%$ (FLIC 0.736--0.762), 0.747--0.833 at 30\% and 0.347--0.500 at 3\%, where only the oracle exceeded 0.500. At $n=3000$ the share of TabICLv2 was 0.013--0.027 at $\pi=3\%$, 0.080--0.127 at 10\% and 0.287--0.387 at 30\%. At 3\% and 10\% its share fell beyond $n=1000$ (at 10\% in S1 it peaked at $n=500$); at 30\% it kept rising. The MCSE of each share was at most 0.022 for the established methods (500 replicates) and at most 0.041 for the TFMs (150 to 200 replicates).

\subsection{Calibration-in-the-large with rare events (Q3)}\label{sec:res-q3}

\begin{figure}[t]
\centering
\includegraphics[width=\textwidth]{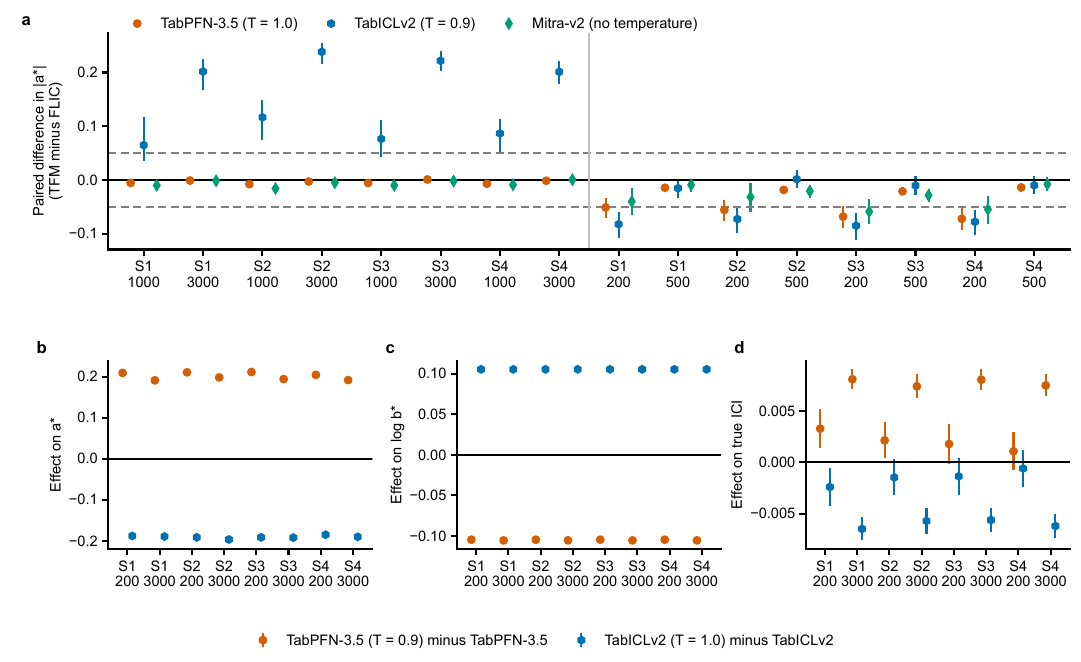}
\caption{\textbf{Calibration-in-the-large with rare events and the effect of the softmax temperature.} \textbf{a}, Paired difference in $|a^*|$ (TFM minus FLIC; Hodges-Lehmann estimate with its 95\% signed-rank interval) in the 16 cells at an event rate of 3\%, the cells with $n\ge1000$ first; dashed lines mark the margin $\pm0.05$, and a negative difference favours the TFM. $T$, softmax temperature; the shipped $T$ of TabPFN-3.5 and TabICLv2 is in the legend; Mitra-v2 has no temperature setting. \textbf{b}--\textbf{d}, Paired effect of the temperature (temperature arm minus main arm) at an event rate of 10\% on $a^*$ (\textbf{b}), $\log b^*$ (\textbf{c}) and the true ICI (\textbf{d}; mean $\pm1.96$ MCSE). TabPFN-3.5 was moved from 1.0 to 0.9 and TabICLv2 from 0.9 to 1.0.}\label{fig:temp}
\end{figure}

TabPFN-3.5 and TabICLv2 form their probabilities after dividing every logit by a softmax temperature, a setting fixed by their developers; a temperature below 1 makes the predictions more extreme. Mitra-v2 has no such setting. At $\pi=3\%$, TabPFN-3.5 (shipped temperature 1.0) was equivalent to FLIC in $|a^*|$ in 12 of 16 cells, better in 1 and inconclusive in 3; Mitra-v2 was equivalent in 12 and inconclusive in 4 (Fig.~\ref{fig:temp}a). Both were equivalent to FLIC in all eight cells with $n\ge1000$ (Hodges-Lehmann estimates $-0.016$ to $0.001$). TabICLv2 (shipped temperature 0.9) was worse than FLIC in 5 cells, inconclusive in 4, equivalent in 4 and better in 3. All four cells at $n=3000$ were worse (estimates 0.201 to 0.239). There its median $a^*$ was 0.250--0.297 (positive: risks too low on average), against $-0.005$ to 0.010 for FLIC, and its mean predicted risk was 0.0062--0.0067 below the true 0.03. At $n=200$ the order reversed: TabICLv2, below its documented minimum of 300 rows, was better than FLIC in 3 cells and TabPFN-3.5 in 1. At $\pi=10\%$ and 30\%, TabPFN-3.5 and Mitra-v2 were equivalent to FLIC in all 32 cells, and TabICLv2 was worse in 8, inconclusive in 5 and equivalent in 19.

The temperature arms estimate this setting's effect in eight cells: S1 to S4 at $\pi=10\%$ with $n$ of 200 and 3,000 (Fig.~\ref{fig:temp}b--d; Supplementary Note~3). Each change shifted $a^*$ by about 0.2, four times its margin. Moving TabPFN-3.5 from 1.0 to 0.9 made it worse in $|a^*|$ than at its shipped setting in 5 of the 8 cells (inconclusive in 3) and raised its true ICI by up to 0.0081. Moving TabICLv2 from 0.9 to 1.0 lowered its true ICI by up to 0.0065, most at $n=3000$. The changes in true ICI were within the margin of 0.01, but at $n=3000$ about half of TabICLv2's mean true ICI.

\subsection{Goodness-of-fit tests on foundation-model predictions (Q4)}\label{sec:res-q4}

The battery has 11 tests of whether observed outcomes agree with predicted risks. Among them are the Hosmer-Lemeshow test, which compares observed and expected events in ten risk groups; EDGE, which projects the residuals of such groups onto a few typical shapes of miscalibration, with 10 groups and with an automatic number of groups; the Cox test that the calibration intercept is 0 and the slope 1; the GiViTI calibration-belt test; and the le Cessie-van Houwelingen test, which also uses the covariates. A test's size is its rejection rate on perfectly calibrated predictions, which should be near 0.05 at the 0.05 level (labels in Methods).

\begin{figure}[t]
\centering
\includegraphics[width=\textwidth]{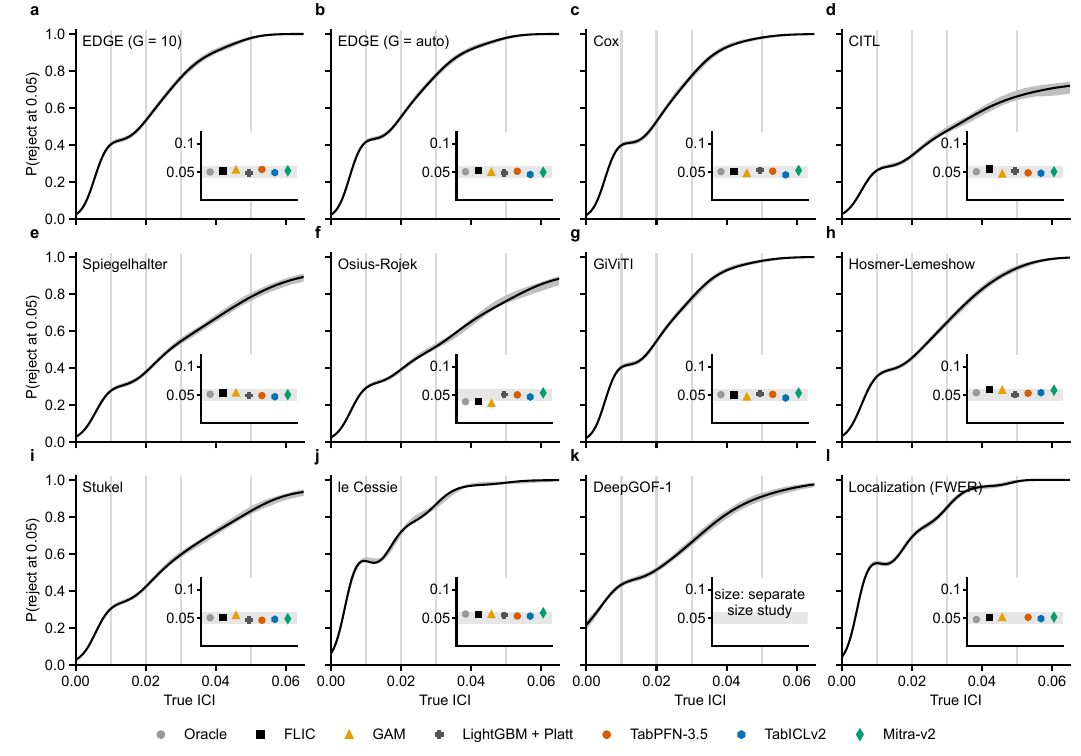}
\caption{\textbf{Detection and size of the goodness-of-fit tests.} Each panel shows the probability that a test rejects at the 0.05 level on the observed outcomes, against the true ICI, from a logistic generalized additive model pooled over models, scenarios (with evaluation sets of 1,000 and 3,334 rows) and replicates, with a 95\% band from a bootstrap that resamples whole replicates. Vertical lines mark true ICI values of 0.01, 0.02, 0.03 and 0.05. Insets show the pooled size of the test on outcomes drawn from each model's own predicted risks ($\pm1.96$ MCSE) against the band $[0.04, 0.06]$; for the oracle, whose observed outcomes are themselves drawn from the true risk, both its observed and its redrawn outcomes are null outcomes. $G$, number of risk groups of EDGE (10, or chosen automatically); CITL, calibration-in-the-large test. DeepGOF-1 is a pretrained neural goodness-of-fit test; its size comes from a separate size study. Panel \textbf{l} is the localization procedure, which names the part of the misfit (overall level, slope, curve shape or covariate subgroup) with the familywise error rate (FWER) controlled at 0.05; it shows the rate of naming any part.}\label{fig:detect}
\end{figure}

\paragraph{Size.} On outcomes drawn from each model's own predicted risks (calibrated by construction), the reference distributions of the tests were accurate. TabPFN-3.5 held its size on all 11 tests; TabICLv2 and Mitra-v2 on 10 of 11 each. For both, the exception was the Hosmer-Lemeshow test, labelled uneven: its pooled size was inside $[0.04, 0.06]$ (0.054 and 0.058, MCSE 0.002) but only 93.1\% and 93.0\% of scenarios lay inside the band. Across the three TFMs and all tests, the pooled size lay between 0.045 and 0.059 (Fig.~\ref{fig:detect}, insets). On the temperature arms, five test--arm pairs were labelled liberal (0.061 to 0.063), though within Monte Carlo error of 0.05 (Supplementary Table~S9). On outcomes drawn from the true risk (58,000 null sets for most tests), 8 tests held their size, the Hosmer-Lemeshow and Stukel tests were uneven, and the Osius-Rojek test was conservative (0.038, MCSE 0.0008). In the random-design tier, every region's size band overlapped $[0.04, 0.06]$, but a typical band ran from about 0.005 to 0.29, so this check could detect only gross distortion.

\paragraph{Detection.} Detection curves are pooled over models, scenarios and evaluation-set sizes (Fig.~\ref{fig:detect}). At a true ICI of 0.01, 0.02, 0.03 and 0.05, EDGE with 10 groups rejected more often than the Hosmer-Lemeshow test by 0.049, 0.078, 0.117 and 0.041 (at 0.01, 0.403 against 0.354). Every 95\% interval lay above the margin of 0.02 (lowest limit 0.031, at 0.05), so EDGE detects more often at all four points. The le Cessie-van Houwelingen test detected miscalibration most often at true ICI 0.01 to 0.03 (0.562 at 0.01). Among tests that use only outcomes and predictions, the Cox test, GiViTI and EDGE (both forms) had almost the same curves (0.400 to 0.417 at 0.01, 0.978 to 0.980 at 0.05). The calibration-in-the-large, Spiegelhalter, Osius-Rojek and Stukel tests detected least often (calibration-in-the-large: 0.663 at 0.05).

\subsection{Recalibration with a local sample (Q5)}\label{sec:res-q5}

Recalibration refits a model's predictions on a local sample from the target population. Logistic recalibration refits the intercept and slope of a logistic regression of the outcome on the logit of the prediction; intercept recalibration refits the intercept only. After logistic recalibration on 500 local rows, each TFM was equivalent to FLIC after the same recalibration in every scenario where it ran (paired differences in true ICI $-0.0076$ to 0.0023). Against the best arm in each core scenario (lowest mean true ICI, recalibrated or as fitted; Methods), each recalibrated TFM was within 0.005 in 30 of the 48 scenarios, unclear in 10 and more than 0.005 above in 8. These 8 were the scenarios with $n=3000$ at $\pi=10\%$ and 30\%, where the best arm was used as fitted, without recalibration (ridge regression in five). At $n=3000$, logistic recalibration on 500 rows raised the true ICI of FLIC, TabPFN-3.5 and Mitra-v2 by 0.004 to 0.011 and that of TabICLv2 by up to 0.005.

\begin{figure}[t]
\centering
\includegraphics[width=\textwidth]{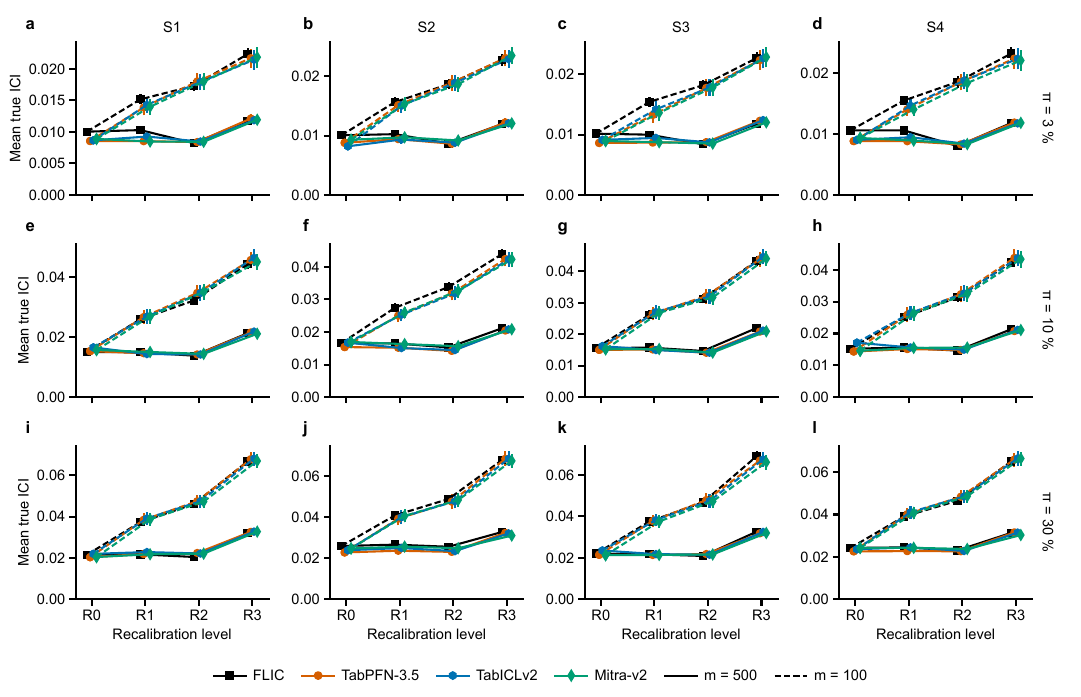}
\caption{\textbf{Recalibration with a local sample.} Mean true ICI at $n=500$ after each level of the ladder: R0, none; R1, intercept only; R2, logistic recalibration; R3, isotonic regression; fitted on a local sample of 500 (solid) or 100 (dashed) rows, for FLIC and the three TFMs. Rows are event rates and columns outcome shapes. Points are means $\pm1.96$ MCSE; with 100 rows, only replicates in which a fit was possible (both outcome classes present) are included. The same figure at $n=200$, 1,000 and 3,000 is in Supplementary Information.}\label{fig:ladder}
\end{figure}

The recalibration ladder adds isotonic regression, a flexible monotone fit, and a local sample of 100 rows (Fig.~\ref{fig:ladder}). With 500 rows, intercept and logistic recalibration kept or slightly lowered the true ICI at $n=500$, and gave large gains only where the unrecalibrated model was poor (S1, $\pi=3\%$, $n=200$: FLIC from 0.0190 to 0.0084). At $n=500$, every level fitted on 100 rows made calibration worse (S1, $\pi=10\%$: TabPFN-3.5 from 0.0153 to 0.0346 after logistic recalibration). Isotonic regression was the worst level with local samples of both 100 and 500 rows.

\paragraph{The test-guided rule.} The rule recalibrates a model logistically when the Cox test rejects, and raises a flag when EDGE still rejects after recalibration. A replicate counts as truly insufficient when its true ICI after recalibration exceeds 0.015 (0.0045 at $\pi=3\%$). The positive predictive value of a flag was 0.864 to 0.883 across the three TFMs and FLIC (lowest 95\% limit 0.849; 2,015 to 2,575 flagged replicates per arm). The negative predictive value was lower. Among replicates where the Cox test rejected but no flag was raised, only 0.472 to 0.522 were adequately calibrated after recalibration.

\subsection{Transfer to randomly drawn designs (Q6)}\label{sec:res-q6}

\begin{figure}[t]
\centering
\includegraphics[width=\textwidth]{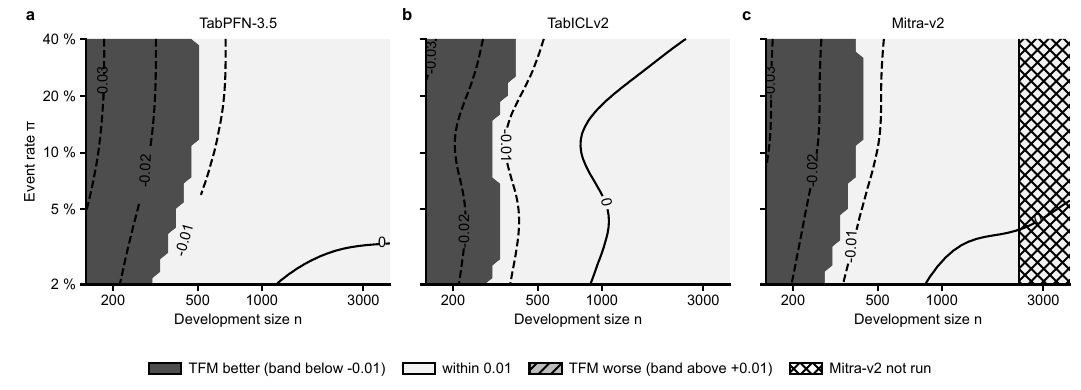}
\caption{\textbf{Where each foundation model is better or worse calibrated than FLIC across random designs.} Meta-model prediction of the paired difference in true ICI (TFM minus FLIC) over development size $n$ and event rate $\pi$, with every other design factor at its median. Dark grey: the 95\% simultaneous band lies below $-0.01$ (TFM better). Hatched light grey: the band lies above $+0.01$ (TFM worse; no such region occurred). White: within 0.01. Dashed and solid lines are contours of the predicted difference. Cross-hatched: Mitra-v2 not run because it does not fit in 6\,GB of graphics memory. Based on 1,000 random designs with one replicate each.}\label{fig:map}
\end{figure}

A generalized additive model (the meta-model) relates each TFM's paired difference from FLIC in the random designs to the design factors. A core scenario agrees when the meta-model's prediction at its settings lies within 1.96 combined standard errors of its own estimate. The grid transfers when at least 40 of the 48 core scenarios agree, and transfers in part when 30 to 39 agree. It did not transfer for any TFM: 12 core scenarios agreed for TabPFN-3.5 and TabICLv2 and 15 for Mitra-v2. None agreed at $n=200$, where the meta-model predicted a larger TFM advantage than the core design. At $n=3000$ it predicted small TFM losses (0.006 to 0.012 in true ICI on average) where the core design showed none (Supplementary Note~3). As the protocol specifies, the random-tier map is therefore the general result. On the map (Fig.~\ref{fig:map}), each TFM was better than FLIC by more than 0.01 in 24.8\% (TabICLv2, 248 of 999 designs with development events) to 38.0\% (Mitra-v2, 315 of the 828 designs where it ran) of designs, mostly small designs with many predictors, and worse in none. For the calibration slope, each TFM was better than FLIC by more than $\log1.1$ in 54.9\% to 63.4\% of designs and worse in 3.6\% to 4.7\%.

\subsection{Real-data learning curves and plasmode tier}\label{sec:res-real}

\begin{figure}[t]
\centering
\includegraphics[width=\textwidth]{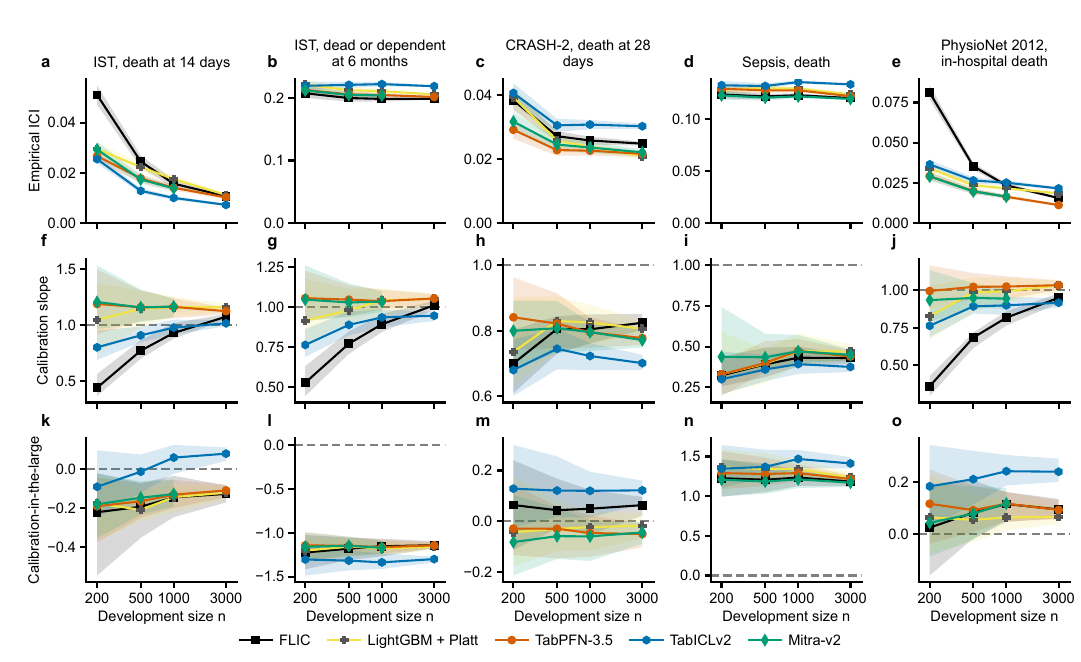}
\caption{\textbf{Calibration of each model on the external real-data sets.} Empirical calibration on the external evaluation set against development size $n$, for FLIC, LightGBM with Platt scaling and the three TFMs, in the International Stroke Trial (IST; two outcomes; developed in the United Kingdom, evaluated in the other countries), CRASH-2 (74 held-out hospitals), the sepsis data (developed in admissions not meeting the Sepsis-3 definition, evaluated in Sepsis-3 admissions) and the PhysioNet 2012 data (sets A and B against set C) (columns). In IST death or dependency (\textbf{b}, \textbf{g}, \textbf{l}) and sepsis (\textbf{d}, \textbf{i}, \textbf{n}), the event rate differs between the two populations. \textbf{a}--\textbf{e}, Mean empirical ICI over 100 development subsamples at each $n$, with 95\% intervals ($\pm1.96$ MCSE). \textbf{f}--\textbf{j}, Median empirical calibration slope with its interquartile range; the dashed line marks 1. \textbf{k}--\textbf{o}, Median calibration-in-the-large (log-odds scale, slope fixed at 1; positive means predicted risks too low) with its interquartile range; the dashed line marks 0. Mitra-v2 was not run at $n=3000$ in IST and PhysioNet 2012, where it does not fit in graphics memory. CRASH-2 results are aggregates only. The real-data results are exploratory.}\label{fig:real}
\end{figure}

These results are exploratory (Methods). On the external sets, TabPFN-3.5 and Mitra-v2 had a lower empirical ICI than FLIC at small $n$ for death at 14 days in the International Stroke Trial (IST), in CRASH-2 and in PhysioNet 2012, and the gap shrank with $n$ (Fig.~\ref{fig:real}a,c,e); TabICLv2 did so except in CRASH-2. The paired difference for TabPFN-3.5 was $-0.024$ (95\% interval $-0.029$ to $-0.020$) at $n=200$ and $-0.0005$ at $n=3000$ for death at 14 days.

Where the event rate changed between populations (IST death or dependency, 80\% against 55\%; sepsis, 4.9\% against 19\%), every model had a large calibration-in-the-large error at every $n$ (medians $-1.1$ to $-1.5$ and $+1.2$ to $+1.5$; Fig.~\ref{fig:real}l,n). There, TabPFN-3.5 and TabICLv2 had a higher empirical ICI than FLIC at every $n$ (by 0.0018 to 0.0062 and 0.0088 to 0.024), and Mitra-v2 a higher one in IST at $n\ge500$ and a lower one, by at most 0.0011, in sepsis. TabICLv2 was also higher at $n=3000$ in CRASH-2 and PhysioNet 2012; for death at 14 days it was lower at every $n$ in the primary split (Methods). In the IST temporal split, TabICLv2 was higher than FLIC at $n=3000$ for both outcomes, whereas TabPFN-3.5 and Mitra-v2 were lower at small $n$ (Supplementary Table~S18a). In CRASH-2, every model remained over-confident at $n=3000$ (median slopes 0.70 to 0.85).

Logistic recalibration on 500 local rows removed most of the shift error for FLIC and the TFMs (ICI from 0.20--0.22 to 0.014--0.023 for IST death or dependency, and from 0.12--0.14 to 0.025--0.033 in sepsis; Supplementary Table~S18b); 100 rows did less. After it, EDGE still rejected in 37\% to 100\% of subsamples across outcomes. For IST death or dependency, localization, which names the part of the misfit, named the overall risk level in every subsample (Supplementary Table~S18d).

The plasmode tier broadly reproduced the simulation for the TFMs (Supplementary Table~S18c). The median $b^*$ was 0.97 to 1.08 for TabPFN-3.5 and 0.92 to 1.03 for Mitra-v2, but below 1 at every $n$ for TabICLv2 (0.77 to 0.93). The tier was harsher on FLIC. Its median $b^*$ was below 1 at every $n$ in every data set (0.40 to 0.97), and 0.81 to 0.83 in CRASH-2 at $n=3000$, against 0.98 to 0.99 in the simulation. Under exact nulls, rejection rates were 0.048 to 0.060, except for the Osius-Rojek test (0.018 for the oracle, 0.038 for the arms; Supplementary Table~S18e).

\section{Discussion}\label{sec:discussion}

Even 3,000 development patients did not guarantee good calibration for any model in this study, classical or foundation. To our knowledge, this is the first pre-registered study to measure the calibration error of the newest tabular foundation models against the true risk of every patient; the evaluations we found estimated calibration from observed outcomes on real data\cite{wu2026,varghese2026,cui2026,demelocosta2026,zhao2026tabrelibench}. At $n=3000$, only 63\% to 83\% of samples gave a weakly calibrated model with FLIC, TabPFN-3.5 or Mitra-v2 at event rates of 10\% and 30\%, and 35\% to 50\% at 3\%. It holds for correctly specified logistic regression too, so it reflects learning a risk from a finite sample, not one method. The minimum sample sizes of Riley and colleagues (323 to 1,845 in the core scenarios) are set so that the expected calibration slope is about 0.9\cite{riley2020}. That value lies on the edge of the weak class, which requires $|b^*-1|\le0.1$, so many samples of that size are expected to fall outside it. The criteria limit overfitting on average; they do not promise a calibrated model from a given sample, and individual risks carry their own uncertainty\cite{riley2025uncertainty}. A model developed on a few thousand patients should still be checked before use.

Tabular foundation models are not a shortcut to calibration. Their calibration slope was within the pre-registered margins of FLIC in 36 or 37 of the 48 core scenarios. Neither they nor FLIC reached the 80\% rule by $n=3000$ in most cells, and with looser thresholds no foundation model needed a smaller sample than FLIC. Their one consistent advantage was at the smallest samples, where FLIC was over-confident and TabPFN-3.5 and Mitra-v2 had slopes closer to 1; the gain in true ICI reached the margin only at an event rate of 3\% with $n=200$. A foundation model removes some of the over-confidence of a small sample; it does not remove the need for an adequate sample or for validation.

TabICLv2 was worse at its shipped settings. It was over-confident in 4 of the 8 main cells of Q1, its true ICI reached a plateau beyond $n=1000$, its share of weakly calibrated samples fell as $n$ grew at event rates of 3\% and 10\%, and at 3\% with $n=3000$ it was worse than FLIC in calibration-in-the-large in every shape. Much of this appears to come from its shipped softmax temperature of 0.9: its median slope settled near 0.9, again the edge of the weak class, whereas at a temperature of 1.0 (event rate 10\%, $n=3000$) its slope was close to 1 and 62\% to 71\% of samples were weakly calibrated, against 8\% to 13\% at the shipped setting (Supplementary Note~3). Users should therefore record such shipped settings and check calibration-in-the-large on their own data.

Logistic recalibration on 500 local rows made every foundation model equivalent to FLIC after the same recalibration. With 100 rows or isotonic regression it usually made calibration worse, and with a large development sample from the target population it also did harm: at $n=3000$ it raised the true ICI of FLIC, TabPFN-3.5 and Mitra-v2 by 0.004 to 0.011.
The fixed grid did not transfer to the random designs. Across those designs each foundation model was better than FLIC by more than 0.01 in a quarter to two-fifths of the space, mostly in small designs with many predictors, and worse in none, but the meta-model also predicted small losses at $n=3000$ that the grid did not show. Part of the disagreement may come from the meta-model. In a post hoc, exploratory analysis, adding an interaction between $n$ and the number of predictors raised agreement from 12, 12 and 15 to 26, 27 and 29 of the 48 core scenarios, still below the 30 needed for partial transfer, and disagreement remained at $n=500$ and 1,000 (Supplementary Note~3). The rest may come from the grid, whose eight predictors may miss settings where the models differ. Comparison studies should check whether their grid transfers, with a pre-specified test, and report the result whatever it is.

The goodness-of-fit tests can be used on foundation-model predictions: their reference distributions were accurate, with the Hosmer-Lemeshow test uneven on TabICLv2 and Mitra-v2. With outcomes and predictions alone, the Cox test, GiViTI and EDGE detected miscalibration most often, and EDGE more often than the Hosmer-Lemeshow test; with covariates, the le Cessie-van Houwelingen test did. The test-guided rule gave reliable positive flags, but among replicates where the Cox test rejected, about half of those without a flag were still inadequately calibrated, so a missing flag is no reassurance.

On real data (exploratory), the small-sample advantage of TabPFN-3.5 and Mitra-v2 reappeared where the event rate did not shift, but no model was protected from a shift in event rate; local recalibration largely removed it. The results should be used within the following rules. They describe the named model versions at their shipped settings; a new version or setting needs its own check. Mitra-v2 received categorical predictors as integer codes, because its interface has no categorical option, and its results cover only designs that fit in 6\,GB of graphics memory. The simulations validate each model in its development population; for a new population, the real-data results and local recalibration apply. The margins are wide relative to the quantities compared, so ``equivalent'' means ``within these margins''; readers who need a stricter tolerance should use the shares and intervals reported for every cell. ``More than 3,000'' in the minimum-$n$ table means that the required size lies beyond the design.

In summary, the newest tabular foundation models did not deliver calibrated risks with less data than penalized logistic regression; they helped mainly at the smallest samples, and TabICLv2 was worse at its shipped settings. Whatever the model, predictions should be checked with a test that holds its size, such as the Cox test, GiViTI or EDGE, and recalibrated with the logistic form on at least 500 local patients when the development sample is small or the population differs, but not when a large development sample comes from the same population.

\section{Methods}\label{sec:methods}

\subsection{Study design and pre-registration}\label{sec:design}

The study has two parts: a simulation study, in which the true risk of every patient is known, and an analysis of four real clinical data sets. The simulation follows the ADEMP structure of Morris, White and Crowther\cite{morris2019}: aims, data-generating mechanisms, estimands, methods and performance measures. The protocol was registered on OSF Registries with the template for simulation studies of Siepe and colleagues\cite{siepe2024}, which follows the same structure. The registration\cite{osf2026} (\url{https://osf.io/8hmtz}) was made on 8 October 2026, and the main simulation run started after it: the run scripts refused to start until the registration link and time had been recorded. The protocol was not edited after registration. In this paper the word ``pre-registered'' is used only for the analyses that the registration describes.

The protocol is estimation-focused. It states no confirmatory hypothesis and predicts the direction of no result. For each research question it fixes in advance the quantity to estimate, the comparison, the summary with its Monte Carlo standard error, and the rule that decides whether a model is called better than, worse than or equivalent to the reference. Before the registration, a pilot run was made on a separate random-number stream. It informed several design choices, which were made before the main run and are listed with their reasons in Supplementary Note~1. No pilot data enter the main results.

The study compares three tabular foundation models with five established methods for predicting a binary clinical outcome. A tabular foundation model (TFM) is a neural network pretrained on many simulated tables. It reads the development sample of the user as a prompt and returns a probability for each new patient, with no fitting of parameters to the user's data. The focus is on calibration: the agreement between predicted and actual risks\cite{vancalster2019}. A model is well calibrated when, among patients given a predicted risk of 20\%, about 20\% have the event.

\subsection{Data-generating mechanisms}\label{sec:dgm}

\paragraph{Covariates.} The core design has eight covariates. $X_1,\dots,X_6$ are continuous and marginally standard normal, with $\operatorname{corr}(X_1,X_2)=0.5$ and all other pairs independent; $X_7\sim\mathrm{Bernoulli}(0.5)$ and $X_8\sim\mathrm{Bernoulli}(0.3)$.

\paragraph{Outcome models.} Let
\begin{equation}\label{eq:L}
L = 1.0X_1 + 0.8X_2 + 0.6X_3 + 0.4X_4 + 0.2X_5 + 0.0X_6 + 0.5X_7 + 0.3X_8
\end{equation}
be the base linear term, in which $X_6$ is pure noise. Four outcome shapes are used:
\begin{itemize}
\item S1, linear: $\eta = \alpha + \kappa L$ with a logit link;
\item S2, smooth non-linear: $\eta = \alpha + \kappa\{L + \gamma[(X_1^2-1)/\sqrt{2} + (X_3^3-3X_3)/\sqrt{6}]\}$ with $\gamma=0.5$ and a logit link;
\item S3, interactions: $\eta = \alpha + \kappa(L + 0.5X_1X_7 + 0.5X_3X_4)$ with a logit link;
\item S4, asymmetric link: $\eta=\alpha+\kappa L$ with the complementary log-log link.
\end{itemize}
The true risk of a patient is $p^* = 1/\{1+\exp(-\eta)\}$ in S1 to S3 and $p^* = 1-\exp\{-\exp(\eta)\}$ in S4, and the outcome is $Y\sim\mathrm{Bernoulli}(p^*)$. The statistical models below use main effects only. They are therefore correctly specified in S1 (all of them) and S2 (the generalized additive model only), and misspecified in S3 and S4. In S3, patients with $X_7=1$ have a steeper risk gradient in $X_1$ than any main-effects model can represent; this subgroup is the target of the localization analysis.

\paragraph{Signal strength and event rate.} For each shape, the scale factor $\kappa$ was solved numerically on $10^6$ fixed draws so that the oracle area under the receiver operating characteristic curve (AUC), the AUC of the true risk $p^*$, equals 0.80 at an event rate of 10\%. For each shape and each event rate $\pi\in\{3\%,10\%,30\%\}$, the intercept $\alpha$ was then solved on the same draws so that the mean true risk equals $\pi$. Because $\kappa$ is fixed at 10\%, the oracle AUC lies between 0.77 and 0.83 at the other two event rates.

\paragraph{Development size.} The development sample size is $n\in\{200,500,1000,3000\}$. Each scenario is also described by its events per variable, $n\pi/p$ with $p$ the number of model parameters, and by the minimum sample size of Riley and colleagues for developing a prediction model\cite{riley2020}, computed with the R package \texttt{pmsampsize} (expected shrinkage 0.9, Cox-Snell $R^2$ derived from the oracle AUC). The scenario with $n=200$ at $\pi=3\%$ (about six events, 0.75 events per variable) is kept on purpose: it is the regime in which a pretrained model might help most. A failed or degenerate fit there is a result, not a reason to drop the scenario.

\paragraph{Scenarios.} The core factorial crosses 4 shapes, 3 event rates and 4 sizes: 48 scenarios. A stress set adds 10 scenarios at $\pi=10\%$ and $n\in\{500,3000\}$: T1, heavy-tailed covariates ($X_1$ to $X_6$ drawn as Student $t$ on 3 degrees of freedom through a Gaussian copula, scaled to unit variance, shape S1); T2, many noise columns (22 independent standard normal columns added, so $p=30$, shape S1); T3, a six-level categorical covariate in place of $X_8$, with level effects $(0, 0.3, -0.3, 0.6, -0.6, 0.9)$ inside $L$ (shape S3); and T4a and T4b, the S2 shape with $\gamma=0.25$ and $\gamma=1.0$. With S2 at $\gamma=0.5$, T4a and T4b form a three-step severity ladder. Supplementary Table~S1 lists all 58 scenarios with their solved $\kappa$ and $\alpha$, oracle AUC, events per variable and Riley minimum sample size.

\paragraph{Data drawn in each replicate.} Each replicate draws three independent sets: a development set of $n$ rows; an evaluation set of $N_{\mathrm{eval}} = \max(1000, \lceil 100/\pi\rceil)$ rows (3,334 at $\pi=3\%$ and 1,000 otherwise), so that it holds at least 100 expected events\cite{riley2021}; and a recalibration set of 500 rows. A probe set of 500 rows per scenario is drawn once and reused in every replicate, to measure how much the prediction for a fixed patient varies between development samples\cite{rileycollins2023}. Smaller validation samples are the first $\lceil 50/\pi\rceil$ rows of the evaluation set at every event rate, and the first $\lceil 200/\pi\rceil$ rows at $\pi=30\%$ only (at the lower rates this would need more rows than the evaluation set holds); no model is refitted for them.

\paragraph{Random numbers.} Every data set is drawn from a seed built from the scenario, the sample-size index and the replicate number, split into independent child streams for the development, evaluation and recalibration sets. The pilot and the test runs used a different stream, which the main run never touches. A SHA-256 fingerprint of each data set is stored with the replicate and checked before every later step; a mismatch, or two replicates with identical data, stops the run. Procedures in R that draw random numbers receive an explicit seed derived from the scenario, replicate, model and test.

\subsection{The random-design tier}\label{sec:random}

A fixed grid of scenarios can be criticised for depending on the settings chosen. A second tier therefore draws every setting at random. It has $D=1{,}000$ designs, each with one replicate. Design $d$ draws its event rate from a log-uniform distribution on $[0.02, 0.40]$; its development size from a log-uniform distribution on $[150, 4000]$, rounded; the number of predictors uniformly from the integers 4 to 30; an exchangeable predictor correlation uniformly on $[0, 0.6]$; a target oracle AUC uniformly on $[0.65, 0.90]$; three severities, of non-linearity $w_{\mathrm{nl}}$, of interaction $w_{\mathrm{int}}$ and of link asymmetry $w_{\mathrm{link}}$, each uniformly on $[0,1]$; and the fraction of discrete predictors uniformly on $[0,0.5]$ (Supplementary Table~S3).

The predictors come from a latent Gaussian vector. Each discrete predictor is categorical (3 to 6 levels) with probability 0.25 and binary otherwise. The linear term has random coefficients, with each predictor after the second set to pure noise with probability 0.25, and is rescaled to the spread of the core design's $L$. The non-linear term of S2 (at $\gamma=1$) and the interaction term of S3 are added with weights $w_{\mathrm{nl}}$ and $w_{\mathrm{int}}$, and the true risk mixes the logit and complementary log-log links,
\begin{equation}\label{eq:mixlink}
p^* = (1-w_{\mathrm{link}})\,\frac{1}{1+e^{-\eta}} + w_{\mathrm{link}}\{1-\exp(-e^{\eta})\}.
\end{equation}
For each design, $\kappa$ and $\alpha$ were solved on 200,000 calibration draws so that the oracle AUC equals the target and the mean risk equals $\pi$. The evaluation set is sized as in the core design; there is no recalibration or probe set. The tier runs the oracle, the five established methods and the three main TFM arms at their shipped settings. The test layer is the same as in the core design, with 99 Monte Carlo draws.

TabPFN-3.5 runs on the graphics processor when $(n+N_{\mathrm{eval}})\times p \le 80{,}000$ and on the central processor otherwise; the device was fixed by this rule before the run and is recorded for every design. Mitra-v2 needs memory that grows with $n^2p$. Above $n^2p = 8.0\times10^7$ it does not fit in the 6\,GB graphics memory of the study computer, so it is recorded as ``not run'' for those designs. Its context is never cut, and its analyses cover only the region in which it ran.

\subsection{Model arms}\label{sec:arms}

Every arm receives the same development set of a replicate (the TFMs read it as a prompt; the other arms are fitted to it) and predicts the same evaluation, recalibration and probe rows, in a single prediction call per replicate (Mitra-v2 splits that call into batches internally). No arm is tuned on evaluation or recalibration data. The TFMs are used with the settings their developers ship and are not tuned at all; the only exception is the softmax temperature of the two temperature arms below. No correction for class imbalance was applied to any arm. Every comparison between two arms uses the replicates in which both produced usable predictions. The arms, with versions, checkpoints and SHA-256 checksums, are listed in Supplementary Table~S2.

\begin{itemize}
\item \textbf{Oracle.} The true risk $p^*$ itself. It defines a perfectly calibrated model and is the basis of the main size study of the tests.
\item \textbf{Maximum-likelihood logistic regression} (R \texttt{stats::glm}, main effects). Separation is recorded, not hidden.
\item \textbf{Firth logistic regression with intercept correction (FLIC)}, the primary statistical reference. Firth's penalty\cite{firth1993} removes the first-order bias of the coefficients and keeps them finite under separation. The intercept correction of Puhr and colleagues\cite{puhr2017} then refits the intercept so that the mean predicted risk equals the observed event rate. It was fitted with \texttt{logistf::logistf(flic = TRUE)} at the package defaults.
\item \textbf{Ridge logistic regression} (R \texttt{glmnet}\cite{friedman2010}), with the penalty chosen by 10-fold cross-validated deviance, folds stratified on the outcome, over 200 log-spaced values from the largest useful penalty down to $10^{-6}$ times it, with the early path stop switched off. An optimum at either end of the path is flagged.
\item \textbf{Generalized additive model} (R \texttt{mgcv}\cite{wood2011}, restricted maximum likelihood), with a smooth function (basis dimension 5) for each continuous covariate and linear terms for binary and categorical covariates.
\item \textbf{LightGBM}\cite{ke2017} (Python LightGBM 4.7.0), a gradient-boosted tree ensemble. Five-fold stratified cross-validation chooses the number of leaves (7, 15 or 31) and the minimum leaf size (10 or 30), with learning rate 0.05 and up to 2,000 rounds with early stopping after 50; the model is then refitted on all $n$ rows. Its scores are recalibrated by Platt scaling, a logistic regression of the outcome on the logit of the out-of-fold scores. If the Platt slope is zero or negative, the raw scores are kept and the replicate is flagged. The raw scores are also stored as a separate column.
\item \textbf{TabPFN-3.5}\cite{hollmann2025,jaeger2026tabpfn35}, the primary TFM (\texttt{tabpfn} 9.1.0, standard checkpoint of 9 September 2026), at its shipped settings: an automatic number of ensemble members and softmax temperature 1.0. The faster variant of the same model was timed and not used, because the standard checkpoint fitted the compute budget.
\item \textbf{TabICLv2}\cite{qu2025tabicl,qu2026tabiclv2} (\texttt{tabicl} 2.2.0, checkpoint of 12 February 2026), at its shipped settings: 8 ensemble members and softmax temperature 0.9. Its documented minimum is 300 training rows, so its results at $n=200$ are reported and flagged.
\item \textbf{Mitra-v2}\cite{tao2026mitra} (through \texttt{autogluon.tabular} 1.6.3\cite{erickson2020}), used in context only, with no fine-tuning, one ensemble member and the whole development set as context. It keeps its shipped reduced-precision (bfloat16) arithmetic, which the study's graphics card emulates. It is not run at T2 with $n=3000$, where it does not fit in graphics memory.
\end{itemize}

Three further arms are reported beside the main comparison but are not part of it. Two \emph{temperature arms} change only the softmax temperature, which divides every logit before the final probabilities are formed (Mitra-v2 has no such setting): TabPFN-3.5 at 0.9 and TabICLv2 at 1.0, with the same seeds as the main arms, in the eight scenarios S1 to S4 at $\pi=10\%$ and $n\in\{200, 3000\}$. A \emph{legacy arm}, Real-TabPFN-2 (the version 2 classifier in \texttt{tabpfn} 9.1.0, at its shipped settings), links the results to the earlier TabPFN literature, in the 16 core scenarios at $\pi=10\%$.

Binary and categorical covariates are declared categorical to TabPFN and TabICLv2. Mitra-v2 receives them as integer codes, because its interface has no categorical argument. The statistical arms receive dummy codes. Other TFMs were considered and excluded before the run, for reasons fixed in the protocol: a hosted model with no published weights cannot be version-pinned and would require sending patient rows to a third party; a model pretrained on real public tables risks overlap with the real data sets; one model did not fit the available graphics memory; and one supports regression only.

\subsection{Estimands}\label{sec:estimands}

All quantities below are computed on the evaluation set of each replicate. Predictions $\hat p$ are clipped to $[10^{-6}, 1-10^{-6}]$, and the number of predictions within $10^{-6}$ of 0 or 1 is recorded.

Because the simulation knows each patient's true risk, calibration can be measured against the truth rather than against noisy 0/1 outcomes. We call such quantities \emph{oracle estimands}. They answer the question ``how far are these predictions from the real risks?'' directly, without the sampling noise of an outcome-based estimate.

\paragraph{True calibration slope.} The true calibration slope $b^*$ is the slope $b$ in the quasi-binomial logistic model
\begin{equation}\label{eq:slope}
\operatorname{logit} \mathrm{E}(p^*\mid\hat p) = a + b\,\operatorname{logit}\hat p,
\end{equation}
fitted with $p^*$ as the response. A slope of 1 is ideal. A slope above 1 means the predictions are under-confident (too close to the average risk); a slope below 1 means they are over-confident (too extreme)\cite{vancalster2016}.

\paragraph{True calibration-in-the-large.} The true calibration intercept $a^*$ is the intercept of the same model with the slope fixed at 1, that is, with $\operatorname{logit}\hat p$ as an offset. It measures calibration-in-the-large: whether the predictions are, on average, too high ($a^*<0$) or too low ($a^*>0$) on the logit scale\cite{vancalster2016}. The difference between the mean prediction and the mean true risk is reported beside it.

\paragraph{True integrated calibration index.} The integrated calibration index (ICI) of Austin and Steyerberg\cite{austin2019} is the mean absolute distance between a prediction and the smoothed actual risk among patients with that prediction. Here the smoothed curve $c(\cdot)$ is a loess fit\cite{cleveland1979} (degree 2, span 0.75, Gaussian family, no robustness iterations) of the true risk $p^*$ on $\operatorname{logit}\hat p$, and
\begin{equation}\label{eq:ici}
\mathrm{ICI}^* = \frac{1}{N_{\mathrm{eval}}}\sum_{i=1}^{N_{\mathrm{eval}}} \bigl|\hat p_i - c(\hat p_i)\bigr|.
\end{equation}
The true ICI is the primary estimand. A loess curve cannot follow the logistic curve exactly, so the estimator has a small floor even for the oracle; the oracle row is shown in every table as that floor. A log-scale version, which smooths $\operatorname{logit}p^*$ on $\operatorname{logit}\hat p$ and is exactly zero for the oracle, is reported beside it.

\paragraph{Secondary estimands.} These are the median, 90th percentile and maximum of $|\hat p - c(\hat p)|$ (true E50, E90 and E$_{\max}$); the oracle calibration error, the mean of $|\hat p - p^*|$, an error at the level of the individual patient; and the expected AUC of $\hat p$ against $p^*$. Empirical versions, computed from the observed outcomes as a real validation study would, are also reported: the empirical ICI, E50, E90 and E$_{\max}$ (loess of $Y$ on $\operatorname{logit}\hat p$), the empirical calibration slope and calibration-in-the-large, the Brier score, the log-loss, the AUC, and the 10-bin expected calibration error\cite{naeini2015,guo2017} for comparison with the machine-learning literature. On the probe set, the spread of each patient's prediction across replicates measures prediction instability\cite{rileycollins2023}.

\paragraph{Calibration classes.} Each fitted model in each replicate is placed in a level of the calibration hierarchy of Van Calster and colleagues\cite{vancalster2016}. It is \emph{mean}-calibrated if $|\overline{\hat p} - \overline{p^*}|\le 0.01$; \emph{weakly} calibrated if, in addition, $|a^*|\le0.1$ and $|b^*-1|\le0.1$; \emph{moderately} calibrated if, in addition, the true ICI is at most 0.015; and \emph{strongly} calibrated (approximately) if, in addition, the true E90 is at most 0.03. At $\pi=3\%$ the three thresholds on the probability scale are multiplied by 0.3, because absolute thresholds are easy to meet when the mean risk is 0.03; the thresholds on the logit scale are unchanged. The \emph{minimum $n$ for calibration} of an arm, for a given shape, event rate and class, is the smallest $n$ in the design at which the arm reaches that class in at least 80\% of replicates, or ``more than 3,000'' if no $n$ qualifies. The class results are repeated at half and at double every threshold.

\subsection{External-validation goodness-of-fit tests}\label{sec:gof}

A goodness-of-fit test asks whether the observed outcomes are consistent with the predicted risks; a small $P$ value is evidence of miscalibration. In external validation the predictions are frozen: they were produced without the validation data, so nothing is estimated from those data and each test must use the reference distribution for fixed predictions. All tests were run with the R package \texttt{ebrahim.gof}\cite{ebrahimgof}, version 2.10.0. EDGE, DeepGOF-1 and the localization procedure were developed by the author, and \texttt{ebrahim.gof} is the author's own package; the other tests are established methods by other authors, implemented in the same package. At every start, the test layer compared a checksum of the installed code with that of the registered build and refused any other build.

\paragraph{The battery.} The function \texttt{run.all.external()} was applied to the observed outcomes of each evaluation set. It returns the following tests:
\begin{itemize}
\item EDGE\cite{ebrahim2026edge}, a directed grouped test in external mode with 10 risk groups (the primary EDGE row) and with an automatic number of groups;
\item the Cox recalibration test\cite{cox1958,miller1991}, a likelihood-ratio test that the calibration intercept is 0 and the slope is 1 in a logistic regression of the outcome on $\operatorname{logit}\hat p$;
\item the calibration-in-the-large score test, $(O-E)/\{\sum\hat p(1-\hat p)\}^{1/2}$, where $O$ and $E$ are the observed and expected numbers of events;
\item Spiegelhalter's $z$ test\cite{spiegelhalter1986};
\item the Osius-Rojek test\cite{osius1992} in its external form, described below;
\item the GiViTI (Gruppo Italiano per la Valutazione degli Interventi in Terapia Intensiva) calibration-belt test\cite{nattino2014} in its external mode;
\item the Hosmer-Lemeshow test\cite{hosmer1980} in its external form, with 10 equal-frequency risk groups referred to a $\chi^2$ distribution on 10 degrees of freedom, because no parameters are estimated from the validation data;
\item Stukel's test\cite{stukel1988} with its two terms added to the frozen logit, in score form, described below;
\item the le Cessie-van Houwelingen kernel test\cite{lecessie1991} over the covariates, when the evaluation covariates were stored.
\end{itemize}
The function also returns three descriptive rows with no $P$ value: the ratio of observed to expected events, the calibration slope and the AUC.

\paragraph{Osius-Rojek, external form.} The Pearson statistic $X^2 = \sum_i (y_i-\hat p_i)^2/v_i$, with $v_i = \hat p_i(1-\hat p_i)$, is standardized by its exact moments under frozen predictions: each term has mean 1 and variance $(1-2\hat p_i)^2/v_i$, so
\begin{equation}\label{eq:or}
z_{\mathrm{OR}} = \frac{X^2 - N}{\{\sum_i (1-2\hat p_i)^2/v_i\}^{1/2}},
\end{equation}
referred to a standard normal distribution, two-sided. Unlike the version for a model checked on its own data, no part of the variance is removed for fitted coefficients, because none were fitted.

\paragraph{Stukel, score form.} Stukel's test adds two terms to the frozen linear predictor $\ell = \operatorname{logit}\hat p$: $z_a = \tfrac12\ell^2$ for $\ell\ge0$ and $z_b=-\tfrac12\ell^2$ for $\ell<0$ (zero otherwise). Instead of fitting these terms, the package computes the score statistic at the null, $U^\top I^{-}U$ with $U = Z^\top(y-\hat p)$ and $I = Z^\top\operatorname{diag}(v)Z$, where $Z$ holds the two columns scaled to their own information and $I^{-}$ is a rank-aware inverse. It is referred to a $\chi^2$ distribution on 2 degrees of freedom, or on 1 when no prediction reaches 0.5 and the upper term is empty. Before registration, a separate null simulation showed that the likelihood-ratio form of this test rejects too often when only a few records have predicted risks above 0.5, as happens at low event rates, and that the score form holds its nominal level; the package was changed to the score form before the run (Supplementary Note~1).

\paragraph{DeepGOF-1.} DeepGOF-1\cite{ebrahim2026deepgof} is a pretrained neural test that reads a map of standardized residuals over the ranks of two covariates. In external mode (\texttt{deepgof1.external()}), its $P$ value is the rank of the observed score among scores computed on 99 outcome vectors drawn from $\mathrm{Bernoulli}(\hat p)$, so it is an exact Monte Carlo test. It was run on the observed outcomes for FLIC and the three TFMs. It used its ``combined'' reading, which takes the smaller $P$ value of the two-strongest-covariates map and the all-pairs maximum and calibrates that minimum exactly; in designs with more than 12 predictors (scenario T2 and the random designs with $p>12$) it used only the faster ``axes'' reading.

\paragraph{Size of the tests.} The size of a test is its rejection rate when the predictions are perfectly calibrated; a test that holds its size rejects in about 5\% of such cases at the 0.05 level. Size was measured in two ways. First, the oracle arm's predictions are the true risks, so its observed outcomes are an exact null; together with an extra null outcome drawn per replicate, this gives 1,000 null evaluation sets per scenario over 500 replicates. Second, for every arm, an exact null outcome $Y^*\sim\mathrm{Bernoulli}(\hat p)$ was drawn from the arm's own seed stream. Under $Y^*$ the arm's predictions are calibrated by construction, so the rejection rate of each test on $Y^*$ is its size on that arm's own distribution of predictions. This matters because TFMs may produce prediction distributions unlike those of logistic regression. The same battery was also run on the smaller validation subsamples, on both the observed outcome and $Y^*$, to describe size and detection against validation size.

\subsection{Localization}\label{sec:localize}

A test that rejects says that the predictions are miscalibrated, but not how. The function \texttt{localize.external()}\cite{ebrahim2026localize} splits the misfit of frozen predictions into four parts that follow the calibration hierarchy: INTERCEPT (risks too high or too low overall), SLOPE (risks too extreme or too modest), LINK (bends in the curve from the score to the risk) and COV (misfit among patients who share the same predicted risk, such as a missed interaction, which no recalibration can repair). Each part is tested by a Cauchy combination\cite{liu2020} of score tests whose bases are confined to that part and made orthogonal to the parts before it. A part is named only when every intersection hypothesis that contains it is rejected (closed testing\cite{marcus1976}), so the probability of naming any part whose misfit is absent is at most 0.05 (the familywise error rate). The intersections are referred to 999 Monte Carlo draws of $Y\sim\mathrm{Bernoulli}(\hat p)$, with a seed per replicate. Localization was run for the oracle, FLIC, the generalized additive model, the TFMs and the temperature arms in every scenario where the covariates were stored; the primary analysis is in S3 and T3, where the true source of misfit is the missed interaction (COV). The oracle rows give the familywise error rate. In T3 the six-level factor enters as five dummy columns.

\subsection{Recalibration}\label{sec:recal}

Recalibration adjusts a model's predictions with a small local sample from the target population\cite{steyerberg2004}. For FLIC and the three TFMs, a local sample of $m\in\{100,500\}$ rows from the recalibration set was used at four levels: R0, none; R1, intercept only (a logistic model with $\operatorname{logit}\hat p$ as offset); R2, logistic recalibration (intercept and slope on $\operatorname{logit}\hat p$); and R3, isotonic regression\cite{zadrozny2002} (pool-adjacent violators). After each level, the oracle estimands and the test battery on the observed outcomes were computed on the evaluation set. A level whose local sample contains only one outcome class is recorded as failed. After R2 on 500 rows, each TFM is compared with FLIC after the same recalibration (with the true-ICI margin, 0.01 or 0.003 at $\pi=3\%$) and with the best arm of each core scenario: the arm with the lowest mean true ICI among FLIC and the TFMs after this recalibration and maximum-likelihood logistic regression, ridge regression, the generalized additive model and LightGBM with Platt scaling as fitted. By a registered rule, a TFM is within 0.005 of the best arm when the upper 95\% limit of the paired difference is at most 0.005, more than 0.005 above it when the lower limit exceeds 0.005, and unclear otherwise.

A test-guided rule was evaluated, not assumed: if the Cox recalibration test rejects at the 0.05 level on the unrecalibrated predictions, apply R2 fitted on the 500-row local sample; if EDGE with 10 groups still rejects at 0.05 after R2, flag ``recalibration insufficient''. A replicate is truly insufficient when its true ICI after R2 exceeds 0.015 (0.0045 at $\pi=3\%$). The positive and negative predictive values of the flag are estimated against this truth.

\subsection{Performance measures and reporting rules}\label{sec:performance}

\paragraph{Replicates and Monte Carlo error.} A simulation estimate varies with the random numbers used; its Monte Carlo standard error (MCSE) is the standard deviation of that estimate across hypothetical repetitions of the whole simulation\cite{morris2019,koehler2009}. The MCSE of a mean is the standard deviation across replicates divided by $\sqrt R$, and that of a proportion is $\{p(1-p)/R\}^{1/2}$, where $R$ is the number of replicates. For medians, a distribution-free 95\% confidence interval from order statistics and a bootstrap MCSE are given. The replicate numbers were fixed in advance from the pilot's variances. The TFMs have $R=200$ replicates at $n\le1000$ and $R=150$ at $n=3000$; the oracle and the established methods have $R=500$; the legacy and temperature arms have $R=100$. The full test battery runs on every arm in replicates 1 to $R$ and on the oracle alone in the remaining replicates up to 500. No replicate was redrawn for any reason, including a development set with no events, and the replicate counts were not changed after any main-run result was seen.

\paragraph{Summaries.} For each scenario and arm, the mean, median, interquartile range and 10th and 90th percentiles of each estimand are reported. For $a^*$ and $b^*$, medians are the main summaries, because a single separated fit can give an extreme value. Every comparison is paired: the TFM minus FLIC on the same replicates, using only the replicates in which both produced usable predictions. For differences in true ICI, the summary is the mean paired difference with MCSE and the 95\% interval mean $\pm 1.96$ MCSE. For differences in $|\log b^*|$ and $|a^*|$, the summary is the Hodges-Lehmann estimate\cite{hodges1963} of the paired differences (the median of all pairwise averages, a robust estimate of the centre), with its distribution-free 95\% interval from the signed-rank statistic and a bootstrap MCSE. A replicate with $b^*\le0$ or no finite $b^*$ counts as the largest possible value of $|\log b^*|$.

\paragraph{Fixed margins and labels.} Each paired difference, in which a negative value favours the TFM, is labelled \emph{better} when its whole 95\% interval lies below $-\delta$, \emph{worse} when it lies above $+\delta$, \emph{equivalent} when it lies inside $(-\delta,+\delta)$, and \emph{inconclusive} otherwise. The margins $\delta$ were fixed before the run: 0.01 for the true ICI (0.003 at $\pi=3\%$), $\log 1.1$ for $|\log b^*|$, and 0.05 for $|a^*|$. For the direction of miscalibration, an arm is called under-confident when the 95\% interval of its median $b^*$ lies above 1.1, over-confident when it lies below 0.9, close to 1 when it lies inside $[0.9, 1.1]$, and unclear otherwise.

For the tests, a test is said to hold its size on an arm when its pooled rejection rate under the null lies in $[0.04, 0.06]$ and its per-scenario rate lies inside the 99\% Monte Carlo band $0.05\pm2.576$ MCSE in at least 95\% of scenarios; it is called liberal above 0.06, conservative below 0.04, and uneven otherwise. Detection is the rejection rate on the observed outcomes. For each test, the probability of rejection is modelled against the true ICI with a logistic generalized additive model (\texttt{mgcv}, a smooth of the true ICI with basis dimension 10), with a cluster bootstrap over replicates (500 resamples) for uncertainty. The main comparison, EDGE against Hosmer-Lemeshow at true ICI values of 0.01, 0.02, 0.03 and 0.05, uses a margin of 2 percentage points. Every table or figure that shows a test's detection also shows its size.

\paragraph{\texorpdfstring{No $P$ value decides a label.}{No P value decides a label.}} The labels come from intervals and fixed margins, not from significance tests. The 95\% intervals are not adjusted for multiplicity; instead, each question reports how many cells fall in each label, so that the reader can weigh the number of comparisons. No statement of ``no difference'', ``equivalent'' or ``holds its size'' is made without the MCSE or interval of the quantity, and a cell at a floor or ceiling is reported as uninformative for that comparison. Each label is reported, whatever it is, in a summary-of-findings table with one row per research question. If an arm fails in more than 5\% of the replicates of a scenario, that scenario is flagged and the arm's results there are reported but not used in the labels.

\paragraph{Failures and degenerate fits.} A fit counts as failed when it raises an error or returns a missing, non-finite or out-of-range prediction; a development set with a single outcome class is a failure for every arm except the oracle. A fit that ran is labelled degenerate when more than 100 of its evaluation predictions lie within $10^{-6}$ of 0 or 1, or when maximum-likelihood logistic regression shows separation. Degenerate fits are the arm's real output: they are scored, tested and kept in every comparison, and they are counted in every table. Known few-event failures (ridge needs at least 3 development events and LightGBM at least 2) are reported as results.

\paragraph{Analysis of the random-design tier.} For each TFM and design, the paired differences against FLIC in true ICI and in $|\log b^*|$ (capped at $\log 10$) are modelled with a generalized additive model (\texttt{mgcv}, restricted maximum likelihood, scaled-$t$ family\cite{wood2016} so that a few separated FLIC fits cannot dominate). It has a smooth term for each design factor (log $n$, log $\pi$, $p$, target AUC and the three severities), separately for each TFM, linear terms for correlation and discrete fraction, and a random effect for the design, which pairs the three TFMs that share its data. Simultaneous 95\% bands come from 10,000 draws of the coefficients. Each design's setting is classified as ``TFM better'' when the band of the predicted true-ICI difference lies below $-0.01$, ``TFM worse'' when it lies above $+0.01$, and ``within 0.01'' otherwise. A transfer check then predicts the difference at each of the 48 core scenarios' own settings: the grid is said to transfer for a TFM when at least 40 of the 48 core scenarios agree with the prediction within $1.96\,(\mathrm{SE}_{\mathrm{meta}}^2+\mathrm{MCSE}_{\mathrm{core}}^2)^{1/2}$, to transfer in part when 30 to 39 agree, and not to transfer otherwise. Rejection by each test is modelled in the same way with a binomial family.

\subsection{Real-data analysis}\label{sec:realdata}

Four public data sets were chosen, with predictor lists fixed from the codebooks before any outcome was tabulated and limited to variables available at the time of prediction. Event rates are computed from the files. Supplementary Table~S19 compares each data set with the simulated design space. (i) The International Stroke Trial\cite{ist1997,sandercock2011,istdata}: death within 14 days, and death or dependency at 6 months, from 20 baseline variables; development in the United Kingdom and external validation in all other countries, with a temporal split as sensitivity analysis (patients randomised before 1994 form the development pool, which holds 3,251 and 3,236 complete cases for the two outcomes, below the 3,500 rule; deviation D-05). (ii) CRASH-2\cite{crash2010,crash2data}: death in hospital within 28 days, from the predictors of Perel and colleagues\cite{perel2012} (Glasgow coma score, age, systolic blood pressure, heart rate, respiratory rate, hours since injury, injury type and sex). The protocol fixed a temporal split, but the released data file has no dates, so whole hospitals were held out instead (74 of the 270 hospitals that remain after excluding patients without a recorded outcome, about 30\,\% of patients; deviation D-01). Hospital identifiers were used only to form this split, and no result is reported by hospital. (iii) Sepsis survival minimal clinical records\cite{chicco2020,chiccodata,knoop2017}: death during the admission, from age, sex and septic episode number; development in the admissions not meeting the Sepsis-3 definition\cite{singer2016} and external validation in the Sepsis-3 admissions. (iv) The PhysioNet/Computing in Cardiology Challenge 2012\cite{silva2012,goldberger2000}: in-hospital death, from 27 summaries of the first 48 hours following the variable list of the Simplified Acute Physiology Score\cite{legall1984}; development in sets A and B and validation in the held-out set C. If a development pool holds fewer than 3,500 complete cases, a pre-specified wider pool is used. If CRASH-2 cannot be obtained, the analysis proceeds with the other three data sets, and this is reported.

In every external region, 500 random rows form the local recalibration sample and the rest form the evaluation set. Missing predictors are handled by complete-case analysis in the stroke and trauma data (sepsis data have none); in the PhysioNet data, medians learned on each development subsample are imputed and a missing-value indicator is added for every variable missing in at least 5\% of the development rows, so every arm receives the same matrix. Three registered sensitivity analyses were not run: a random half split of the sepsis data, a split of the PhysioNet data by type of intensive care unit, and keeping incomplete rows or using the TFMs' own handling of missing values (Supplementary Note~2, D-06).

For each data set and outcome, 100 development subsamples are drawn at each $n\in\{200,500,1000,3000\}$ (or up to the pool size) and each is evaluated on the whole external evaluation set, giving learning curves of the empirical calibration slope, calibration-in-the-large, ICI, E50, E90, Brier score, AUC and the test battery. The real data have no oracle arm. The five established methods and the three TFMs are run at their shipped settings; the temperature and legacy arms are not run on the real data (deviation D-05). As in the simulation, Mitra-v2 is recorded as not run where it does not fit in graphics memory (IST and PhysioNet 2012 at $n=3000$), and TabPFN-3.5 runs on the central processor where the graphics card cannot hold it (PhysioNet 2012 at $n=3000$; deviation D-03). Localization and the recalibration ladder (with local samples of 500 and 100 rows) are applied as in the simulation, with age band, sex, region where present and a severity score as localization covariates.

A plasmode tier places a known true risk on real covariates, so that the oracle estimands can be computed on real patients' characteristics with outcomes that no TFM could have seen during pretraining. For IST (death or dependency at 6 months), CRASH-2 and PhysioNet 2012, a main-effects logistic model is fitted once to the real outcome on the full complete-case table. Two terms of fixed size are added on the logit scale: a U-shaped term $0.4(z^2-1)$ in one standardized continuous predictor $z$, and an interaction $0.6\,z_a x_b$ between a standardized continuous predictor $z_a$ and a binary predictor $x_b$: systolic blood pressure and age $\times$ impaired consciousness in IST, systolic blood pressure and Glasgow coma score $\times$ penetrating injury in CRASH-2, and maximum heart rate and age $\times$ mechanical ventilation in PhysioNet 2012. The intercept is then solved so that the mean true risk is 10\% or 30\%, and outcomes are drawn from this true risk. Development subsamples come from the development region and are evaluated on the external evaluation set. The established methods have 200 replicates per $n$; the TFMs have 50 (deviation D-05), so their plasmode summaries carry a larger Monte Carlo error. The plasmode tier reports the oracle estimands, the summaries of Q1 to Q3 and the size of the tests. Because the real-data analysis depends on major deviations (below), its results are labelled exploratory, and they do not enter the labels. They are reported following TRIPOD+AI\cite{collins2024}. The real-data analyses are secondary analyses of anonymised data, used under each provider's data-use terms (see Ethics declarations). No patients or members of the public were involved in the design, conduct or reporting of this study, which uses simulated data and existing anonymised data only.

\subsection{Software and computing}\label{sec:software}

The TFMs and LightGBM ran in Python 3.11.15 with PyTorch 2.13.0 (CUDA 12.6)\cite{paszke2019}, NumPy 2.4.6\cite{harris2020}, scikit-learn 1.9.1\cite{pedregosa2011}, \texttt{tabpfn} 9.1.0, \texttt{tabicl} 2.2.0 and \texttt{autogluon.tabular} 1.6.3. The statistical arms, the tests and the analyses ran in R 4.4.1\cite{rcore2024} with \texttt{ebrahim.gof} 2.10.0, \texttt{logistf} 1.26.1, \texttt{glmnet} 4.1-10, \texttt{mgcv} 1.9-4, \texttt{pmsampsize} 1.1.3, \texttt{givitiR} 1.3 and \texttt{callr} 3.7.6. The two languages exchanged data only through files, each carrying the data fingerprints. Every software version and the SHA-256 checksum of every model checkpoint were checked at the start of the run, and a mismatch stopped it. The computer was a desktop with an AMD Ryzen 9 3900X processor (12 cores), 32\,GB of memory and an NVIDIA GeForce GTX 1660 Ti graphics card with 6\,GB of memory, running Windows 11. TabPFN-3.5 ran on the graphics card for some groups of scenarios and on central-processor workers for others, as fixed in the protocol; on the same replicate the two devices gave predictions that differed by at most $2.6\times10^{-4}$. The device used for every scenario is reported.

Every replicate was written to disk as soon as it was complete, so a stopped process could resume where it stopped. During the main run, the computer stopped three times (a restart for an operating-system update on 9 October and losses of power on 11 and 13 October 2026), stopping all running processes each time; the run was resumed with the same commands, and no completed replicate was lost (one random design being written at the second power loss was regenerated from its own seed). Separately, in 22 replicates of one scenario (T4a with $n=3000$) the statistical arms were recorded as failed because the operating system failed to start the R process, not because a model failed. As the protocol allows for a crashed process, these 22 replicates were re-run for the missing arms only, and all 22 completed. These and the other run incidents, listed in Supplementary Note~2, changed no data set, setting or analysis.

\subsection{Computing cost}\label{sec:cost}

The time of every fit and every test was stored in the run records (Supplementary Table~S20). On the desktop computer described above, in the core scenarios with $n=3000$, a FLIC fit took a median of 0.07\,s on one processor thread, including the prediction of 2,000 rows. TabPFN-3.5 took 58\,s on the graphics card (scenarios at a 3\% event rate) and 167\,s on six processor threads (the other scenarios); TabICLv2 took 9.5\,s and Mitra-v2 12\,s on the graphics card. Over the 58 scenarios, the established methods used 38 processor hours, the three TFMs 68 graphics-card hours and 194 processor-worker hours, and the test layer 342 single-thread hours. With all streams running at once, the main run, including the random-design tier, took 6.2 days, and the real-data run 12.2 days; both times include the stops of the computer.

\subsection{Deviations from the protocol}\label{sec:deviations}

Every change made after registration is recorded in a dated deviations log with its reason, whether any main-run result had been seen, and its expected effect; the log is given in Supplementary Note~2, together with the run incidents, which the protocol does not count as deviations. No deviation concerns the simulation's analyses. Five are major under the protocol, because they change a real-data split, arm set or replicate count, or drop a registered analysis, and the real-data results that depend on them are therefore labelled exploratory. D-01: in CRASH-2, whole hospitals were held out instead of a split by date, because the released file has no dates. D-03: on the real data, Mitra-v2 is not run where the graphics card cannot hold the model (IST and PhysioNet 2012 at $n=3000$), and TabPFN-3.5 runs on the central processor there (PhysioNet 2012 at $n=3000$). D-05: the temperature and legacy arms are not run on the real data, and the TFMs have 50 instead of 200 plasmode replicates, to fit the available computing time; the IST temporal split is run with the main arms only, on development pools of 3,251 (death at 14 days) and 3,236 (death or dependency) complete cases, below the pre-specified 3,500. No real-data calibration result had been seen when D-01, D-03 or D-05 was decided. D-06: the three registered sensitivity analyses named above were not run, for lack of computing time on the single graphics card; this was decided on 28 October 2026, after the main real-data results had been seen. D-07: localization was not run on the sepsis data, for which the protocol names no severity score, and the ten planned main-text display items were adapted to the journal's limit of eight, with the others in the Supplementary Information; this was also recorded on 28 October 2026, after the main real-data results had been seen. The other deviations are minor: a correction of the registered licence statement for CRASH-2 and implementation details that the protocol leaves open.

\subsection{Use of artificial intelligence tools}\label{sec:llm}

Artificial intelligence tools were used to assist with editing the wording of the manuscript and with checking computer code. The author reviewed and verified all assisted output and takes full responsibility for the content of the publication.

\backmatter

\bmhead{Data availability}
The simulated data are generated by the code described in the Code availability statement; every data set can be rebuilt exactly from its seed and checked against its stored SHA-256 fingerprint. The per-replicate predictions and test results of the simulation, with a loader and a data dictionary, are deposited on Zenodo under a CC BY 4.0 licence (doi:10.5281/zenodo.23737416). The real data are available from their providers under their own terms and are not redistributed here. The International Stroke Trial database is available from Edinburgh DataShare (doi:10.7488/ds/104) under the Open Data Commons Attribution License. The CRASH-2 data are available on request from the London School of Hygiene \& Tropical Medicine Clinical Trials Unit through the freeBIRD repository (doi:10.17037/DATA.10), under the freeBIRD Data Sharing Agreement, which does not allow redistribution; the CRASH-2 results are published with the permission of the trial's Chief Investigators. The Sepsis Survival Minimal Clinical Records are available from the UCI Machine Learning Repository (doi:10.24432/C53C8N) under a CC BY 4.0 licence. The PhysioNet/Computing in Cardiology Challenge 2012 data are available from PhysioNet (\url{https://physionet.org/content/challenge-2012/1.0.0/}) under the Open Data Commons Attribution License. Derived per-patient predictions are deposited only for the data sets whose licence allows it (Sepsis and PhysioNet 2012); for the other two, only aggregate results are deposited. The registered protocol is available at \url{https://osf.io/8hmtz} (doi:10.17605/OSF.IO/8HMTZ).

\bmhead{Code availability}
All code for the simulation, the model arms, the tests and the analyses is available in a public GitHub repository (\url{https://github.com/ebrahimkhaled/tfm-calibration-reproducibility}) under the MIT licence, archived on Zenodo (doi:10.5281/zenodo.23737378). The goodness-of-fit tests use the R package ebrahim.gof, version 2.10.0, which is on CRAN (doi:10.32614/CRAN.package.ebrahim.gof). TabPFN-3.5 weights are distributed by Prior Labs under a licence that does not allow redistribution, so the archive holds no TabPFN weights; it holds a script that downloads them after the user has accepted the licence and checks their SHA-256 checksums. The other model checkpoints are identified by repository, file name, revision and SHA-256 checksum in Supplementary Table~S2.

\bmhead{Acknowledgements}
The author thanks the CRASH-2 Trial Collaborators, whose data were provided by the London School of Hygiene \& Tropical Medicine Clinical Trials Unit through the freeBIRD repository, and the investigators of the International Stroke Trial, the Sepsis Survival Minimal Clinical Records and the PhysioNet/Computing in Cardiology Challenge 2012 for making their data available.

\bmhead{Funding}
The author received no specific funding for this work.

\bmhead{Author contributions}
E.K.E. conceived and designed the study, wrote the protocol, wrote the code, ran the analyses, interpreted the results and wrote the manuscript.

\bmhead{Competing interests}
The author declares no competing interests.

\bmhead{Ethics declarations}
This study is a secondary analysis of anonymised data from completed studies, obtained from their providers under the providers' data-use terms. No new data were collected and no patient was contacted, so no new ethics approval or informed consent was required. All methods were carried out in accordance with relevant guidelines and regulations.

\bmhead{Additional information}
Supplementary Information accompanies this paper. Correspondence and requests for materials should be addressed to E.K.E.

\bibliography{refs}

\end{document}